%% file: main.tex
\documentclass[sigconf,9pt]{acmart}
\renewcommand\footnotetextcopyrightpermission[1]{}
\usepackage{booktabs}
\usepackage{color}
\usepackage{graphicx}
\usepackage{url}
\usepackage{multirow}

\usepackage{xspace}

\usepackage{subfigure}
\usepackage{amsmath,chemformula}

\usepackage{pbox}
\usepackage{threeparttable,booktabs,multicol,multirow}

\setcopyright{rightsretained}
\setcopyright{none}

\begin{document}

\title{LoCI: An Analysis of the Impact of Optical Loss and Crosstalk Noise in Integrated Silicon-Photonic Neural Networks
}

\author{Amin Shafiee$^\dagger$, Sanmitra Banerjee$^*$, Krishnendu Chakrabarty$^*$, Sudeep Pasricha$^\dagger$\\ and Mahdi Nikdast$^\dagger$ }

\affiliation{$\dagger$ Department of Electrical and Computer Engineering, Colorado State University \city{Fort Collins, CO 80523} \country{USA}\\
$^*$Department of Electrical and Computer Engineering, Duke University \city{Durham, NC 27708} \country{USA} }

\input{abs.tex}


\maketitle

\thispagestyle{plain}
\pagestyle{plain}

\input{intro.tex}

\input{Background.tex}

\input{methods.tex}

\input{results.tex}

\input{conclusion.tex}


\bibliographystyle{IEEEtran}
\bibliography{IEEEabrv,ref}
%
%
\end{document}

%% file: abs.tex
\begin{abstract}
Compared to electronic accelerators, integrated silicon-photonic neural networks (SP-NNs) promise higher speed and energy efficiency for emerging artificial-intelligence applications. However, a hitherto overlooked problem in SP-NNs is that the underlying silicon photonic devices suffer from intrinsic optical loss and crosstalk noise, the impact of which accumulates as the network scales up. Leveraging precise device-level models, this paper presents the first comprehensive and systematic optical loss and crosstalk modeling framework for SP-NNs. For an SP-NN case study with two hidden layers and 1380 tunable parameters, we show a catastrophic ~84\% drop in inferencing accuracy due to optical loss and crosstalk noise.
\end{abstract}



\keywords{Integrated Photonic Neural Networks, Optical loss and crosstalk}

%% file: intro.tex
\vspace{-0.08in}
\section{Introduction}
Deep learning has received tremendous interest due to its vastly superior performance across application domains ranging from image recognition to various decision-making problems. However, contemporary electronic deep-learning inference accelerators have shown relatively low energy-efficiency and have been unable to keep up with the performance demands from emerging deep learning applications \cite{SiPh_codesign}. To overcome these bottlenecks, novel accelerators tailored towards artificial intelligence (AI) applications are on the rise, among which integrated silicon-photonic neural networks (SP-NNs) have attracted much attention with a promise of light-speed communication and computation using optical interconnects and silicon photonic devices \cite{sunny2021survey}. For instance, for computationally expensive multiply-and-accumulate operations, optical computing can achieve up to a 1000$\times$ better energy-efficiency footprint compared to electronic accelerators~\cite{totovic2020femtojoule}. \par

\begin{figure}[t]
    \centering
    \includegraphics[width=.47\textwidth]{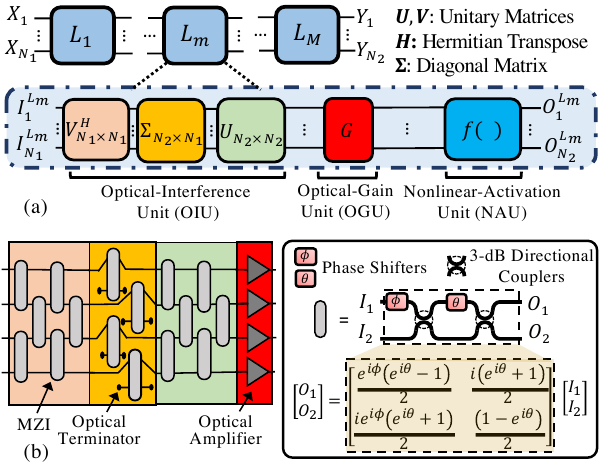}
    \vspace{-0.15in}
    \caption{(a) Overview of a coherent SP-NN with $N_1$ inputs, $N_2$ outputs, and $M$ layers. (b) An optical-interference unit architecture  (left) based on \cite{Clements:16} with $N_1=N_2=$~4, considered as an example, and the underlying 2$\times$2 MZI multiplier (right). }
    \vspace{-0.25in}
    \label{SP-NN_Arch}
\end{figure}

SP-NNs use silicon photonic devices---e.g., Mach--Zehnder interferometers (MZIs)---to realize matrix-vector multiplication with a computational complexity of $O(1)$ \cite{SiPh_codesign}. Among different SP-NN implementations, coherent SP-NNs, which operate on a single wavelength, have an inherent advantage over noncoherent SP-NNs that require power-hungry wavelength-conversion steps and multiple wavelength sources \cite{sunny2021survey}. Fig.~\ref{SP-NN_Arch}(a) presents an overview of a multi-layer coherent SP-NN with $N_1$ inputs, $N_2$ outputs, and $M$ layers. Each layer comprises an optical-interference unit (OIU) implemented using an array of MZIs, connected to a nonlinear-activation unit (NAU) using an optical-gain (amplification) unit (OGU). 

While SP-NNs are promising alternatives to electronically implemented neural networks, several performance roadblocks still need to be addressed. In particular, the underlying silicon photonic devices in SP-NNs suffer from intrinsic optical loss and crosstalk noise due to inevitable device imperfections (e.g., sidewall roughness) and undesired mode couplings \cite{Bahadori:16}. For example, prior work has shown up to 1.5~dB and $-$18~dB for insertion loss and crosstalk, respectively, in 2$\times$2 MZIs \cite{Farhad_4by4}. Note that while the optical loss and crosstalk are small at the device level, they can accumulate as SP-NNs scale up, hence limiting the scalability and degrading the performance of SP-NNs. Even worse, crosstalk noise cannot be filtered in coherent SP-NNs---our focus in this paper---due to the coherence between the noise and victim signals. This necessitates careful analysis of optical loss and crosstalk noise in SP-NNs and their impact on SP-NN performance, which have not been addressed in any prior work. 

The novel contribution of this paper is in developing, to the best of our knowledge, the first comprehensive and systematic optical \underline{Lo}ss and \underline{C}rosstalk modeling framework for \underline{I}ntegrated silicon-photonic neural networks, called LoCI. We develop a realistic device-level MZI compact model to analyze the optical loss from different sources (e.g., propagation loss and metal absorption loss) and the coherent crosstalk noise in the MZI. This model is able to capture the impact of optical phase settings, which represent weight parameters in coherent SP-NNs, on the MZI optical loss and crosstalk performance. Leveraging our accurate device-level models, we present layer- and network-level optical loss and coherent crosstalk models that scale with the number of inputs and layers in coherent SP-NNs. Moreover, LoCI enables an accurate exploration of the power penalty and inferencing accuracy in SP-NNs under optical loss and crosstalk noise. Leveraging LoCI, we also quantify the maximum optical loss acceptable in the underlying devices when specific inferencing accuracy goals must be met with an SP-NN.

Employing LoCI, we analyzed the average and the worst-case optical loss and coherent crosstalk noise in SP-NNs across different numbers of inputs and layers. Our results show considerable degradation in optical signal integrity in the SP-NNs' output layer due to optical loss and crosstalk noise. Considering an example of an SP-NN case study with two hidden layers ($M=$~3) and 16 inputs (i.e., 1380 tunable parameters) with an input optical power of 0~dBm and an OGU with 17~dB optical gain \cite{haq2020micro_SOA_Cband}, we found that the optical loss, optical coherent crosstalk power, and optical power penalty in the output can be as high as 4~dB, 31.7~dBm, and 20~dBm, respectively. Also, we show the inferencing accuracy in this network can drop by 84\% due to optical loss and crosstalk. Increasing the number of inputs from 16 to 32 in the same network, the resulting optical power penalty increases unbearably to as high as 85~dBm. Note that existing work on optical loss and crosstalk analysis in on-chip photonic networks (e.g., \cite{Nikdast_crosstalk}) cannot be applied to SP-NNs as they have unique characteristics that cannot be captured accurately by the models developed for optical interconnects (see Section 2.4). \par

The rest of the paper is organized as follows. Section 2 reviews fundamentals of SP-NNs and prior related work. In Section 3, we present LoCI and the analytical models of optical loss and crosstalk noise at the device, layer, and network level. Section 4 includes simulation results to show the impact of loss and crosstalk on the performance of SP-NNs. Finally, we draw conclusions in Section 5.\vspace{-0.05in}

%% file: Background.tex
\section{Background and Related Work}

\subsection{2$\times$2 MZI Multiplier}
As shown in Fig. 1(b)-right, a 2$\times$2 MZI is the building block of the optical-interference unit in coherent SP-NNs. It consists of two 3-dB directional couplers (DCs), with a nominal splitting ratio of 50:50, and two optical phase shifters ($\theta$ and $\phi$), which are often implemented using microheaters \cite{Farhad_4by4}. Using the phase shifters, one can actively change the phase angle of optical signals traversing the MZI, hence controlling the interference in the output DC and imprinting weight/activation parameters into the electric field amplitude of the optical signals. Accordingly, as shown in \cite{Farhad_4by4} and Fig.~\ref{SP-NN_Arch}(b)-right, an input vector of two optical signals (on $I_1$ and $I_2$) can be coherently multiplied to the transfer matrix of the MZI---defined based on the phase settings on $\theta$ and $\phi$, which represent weight parameters in SP-NNs---to obtain the output vector. We will further discuss the MZI transfer matrix in Section 3.1.\vspace{-0.05in} 

\subsection{Coherent Optical-Interference Unit (OIU)}
Several architectures have been proposed to enable MZI-based linear multipliers (i.e., OIU in Fig.~\ref{SP-NN_Arch}(a)) for deep neural networks \cite{Reck,Clements:16,Shokraneh:20_Diamond}. A fully connected layer $L_{m}$ with $n_m$ neurons performs linear multiplication between an input vector and a weight matrix ($W$) followed by a non-linear activation ($f$). Accordingly, the output of the next layer $L_{m+1}$ can be represented as $O_{m+1} =f_m(W_{m}\times O_m$), where $O_m$ is the output of the previous layer. Using singular value decomposition (SVD), a weight matrix $W$ in layer $L_m$ can be decomposed to $W_m=U_m \Sigma^{m} V_{m}^{H}$, where $U_m$ and $V_{m}^{H}$ are unitary matrices with dimension of $n_m\times n_m$, and $\Sigma^{}_{n_{m}\times n_{m}}$ is a diagonal matrix (see Fig.\ref{SP-NN_Arch}(a)). Here, $V^H$ stands for Hermitian transpose of $V$. Employing the Clements' method in \cite{Clements:16}, $U_m$ and $V_{m}^{H}$ can be mapped into an array of cascaded MZIs (see Fig. \ref{SP-NN_Arch}(b)-left) by adjusting the phase settings on each MZI. The diagonal matrix ($\Sigma^{m}_{n_{m}\times n_{m}}$) can be realized by MZIs with one input and one output being terminated, as shown in Fig.\ref{SP-NN_Arch}(b)-left. Based on~\cite{Clements:16}, the number of MZIs required to implement an $N_1\times N_1$ unitary and an $N_2\times N_1$ diagonal matrix is $\frac{N_1(N_1-1)}{2}$ and $\min(N_1, N_2)$, respectively.\vspace{-0.05in} 

%


\subsection{Optical Loss and Crosstalk Noise}\label{sec::loss_xtalk}
Silicon photonic devices intrinsically suffer from optical loss and crosstalk noise. For example, an optical signal traversing an MZI experiences optical loss through the DCs (e.g., 0.1--0.4~dB \cite{Bahadori:16}), absorption loss due to microheaters' metal planes in proximity (e.g., 0.1--0.3~dB \cite{ding2016broadband}), and propagation loss in the waveguides (e.g., 1--4~dB/cm \cite{Bahadori:16}). Optical crosstalk noise is another limiting factor in silicon photonic networks \cite{Xtalk_pernalty_meisam}. Optical crosstalk is a result of undesired mode coupling among signals of the same wavelength (coherent crosstalk) or different wavelengths (incoherent crosstalk). In coherent SP-NNs with a single wavelength, part of the signal on the same wavelength may leak through a device and experiences a different delay (phase), which is common in coherent networks with cascaded MZIs. Such leaked signals will interfere with the victim signal (see Fig.~\ref{MZI_schematic}) at the output as coherent in-band crosstalk noise. Note that the coherent in-band crosstalk noise is more critical than the incoherent out-of-band noise (exists in noncoherent SP-NNs) due to its coherent nature that makes its filtering impossible.\vspace{-0.09in} 
\subsection{Related Prior Work}\label{sec::related_work}
While several high-performance AI accelerators based on coherent SP-NNs have been recently proposed \cite{PourFard:20,sunny2021survey,Shokraneh:20_Diamond}, \cite{banerjee2021modeling} showed that the inferencing accuracy of SP-NNs can drop by up to 70\% due to fabrication-process variations and thermal crosstalk. In addition to these variations, the work in \cite{Shokraneh:20_Diamond} explored the impact of optical loss non-uniformity among MZIs and showed SP-NN performance degradation. As the size and complexity of emerging SP-NNs increase to handle more complex tasks, the total insertion loss in the network increases as well. This necessitates the use of power-hungry optical amplification devices \cite{haq2020micro_SOA_Cband} and higher laser power at the input. Uncertainties due to fabrication-process variations---the analysis of which is beyond the scope of this paper---in the two DCs in an MZI can degrade the extinction ratio (ER) of the device which, in turn, will increase the loss and crosstalk in the output \cite{De_marinis_app11136232}. Yet, there is no prior work that analyzes the impact of optical loss and crosstalk noise in SP-NNs. While the use of silicon nitride platform can help reduce the loss \cite{De_marinis_app11136232}, the performance degradation due to coherent crosstalk in SP-NNs still remains unaddressed.   \par

Unlike in SP-NNs, optical loss and crosstalk noise have been widely studied in chip-scale Datacom photonic networks (e.g., \cite{Nikdast_crosstalk} and \cite{app10238688_Xtalk}), showing signal integrity degradation and scalablity constraints in these networks due to optical loss and crosstalk noise. Unfortunately, the existing work on optical loss and crosstalk analysis in such networks cannot be applied to SP-NNs as the function, and hence optical loss and crosstalk noise characteristics of silicon photonic devices for optical-domain computation in SP-NNs are different. For example, a 2$\times$2 MZI switching cell, whose structure is similar to the one in Fig.~\ref{MZI_schematic} but without $\phi$, in an optical switch fabric can only assume two functional states based on $\theta$ for optical loss and crosstalk analysis: the Cross-state, where $\theta=0$ and $I_1\rightarrow O_2$ and $I_2\rightarrow O_1$, and the Bar-state, where $\theta=\pi$ and $I_1\rightarrow O_1$ and $I_2\rightarrow O_2$. However, in coherent SP-NNs, $\theta$, which determines the MZI state, can assume any value between 0 and $\pi$ (0$\leq\theta\leq\pi$). The analysis of optical loss and crosstalk in SP-NNs should therefore account for various phase settings in the underlying MZI devices. 

In contrast to prior work, this paper presents the first systematic modeling framework for the optical loss and coherent crosstalk noise from MZI device-level to SP-NN network-level in coherent SP-NNs. Moreover, the developed models at the network level are malleable; the number of inputs and layers in SP-NNs can be varied to evaluate the average and the worst-case optical loss and crosstalk in the network and explore SP-NN power penalty and scalability. 

%% file: methods.tex
\section{Optical Loss and Crosstalk Noise Analysis in Coherent SP-NN\lowercase{s}}\label{sec::LoCI}
\subsection{Device-Level Compact Models}\label{sec::device-compact-models}
\begin{figure}[t]
    \centering
    \includegraphics[width=.45\textwidth]{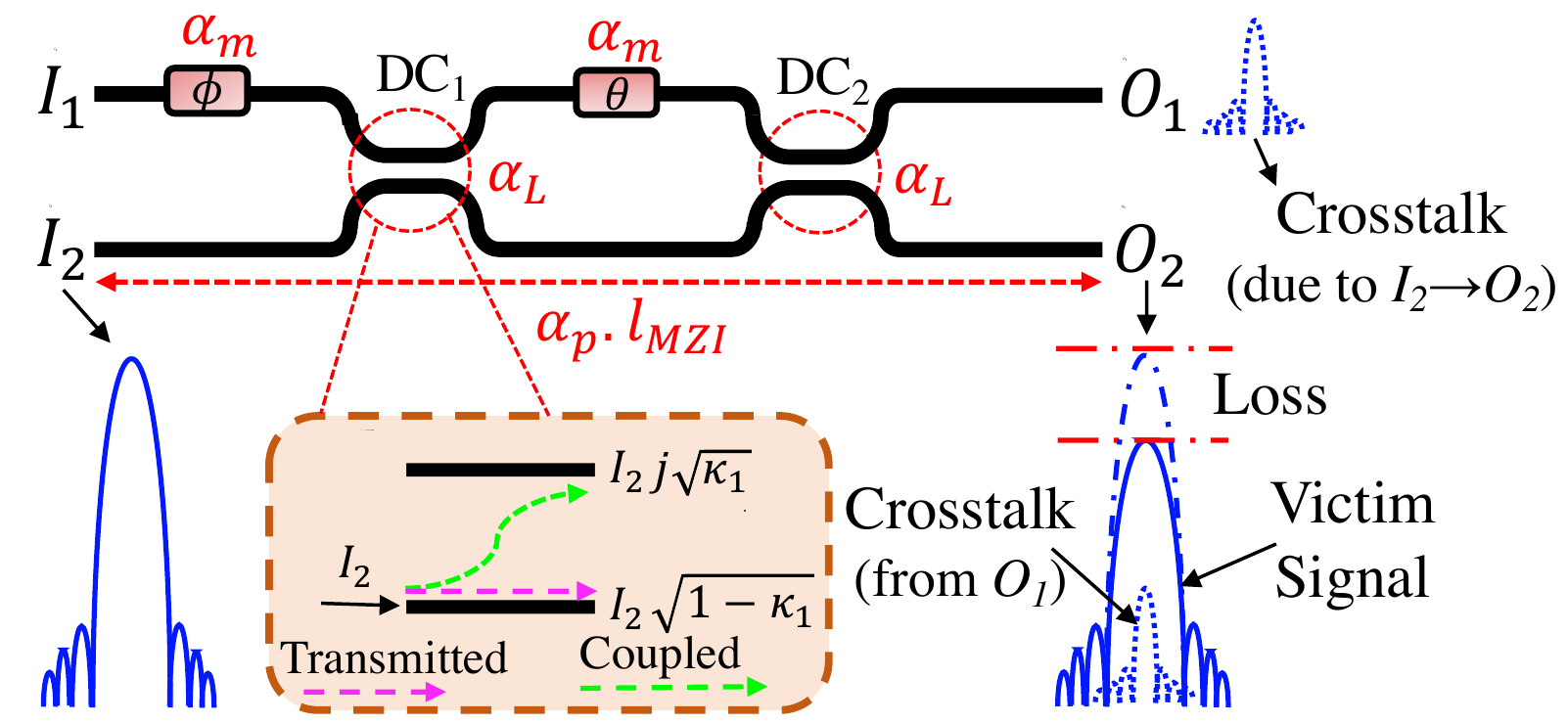}
    \vspace{-0.1in}
    \caption{Schematic of a 2$\times$2 MZI multiplier with different sources of optical loss and crosstalk noise (see Table \ref{table_1}). Here, $I_2\rightarrow O_2$ is shown as an example with $\theta=\pi$ ($l_{MZI}$: MZI length).}
    \vspace{-0.2in}
    \label{MZI_schematic}
   \end{figure}

Fig.~\ref{MZI_schematic} shows a 2$\times$2 MZI structure in coherent SP-NNs. As discussed in Section \ref{sec::loss_xtalk}, the main sources of optical loss in the MZI are the DC loss ($\alpha_L$), the metal absorption loss ($\alpha_m$) through the phase shifters $\phi$ and $\theta$, and the propagation loss ($\alpha_p$) in the waveguides. In DCs (see Fig.~\ref{MZI_schematic}), a fraction (determined by cross-over coupling coefficient $\kappa$) of the optical signal in an input waveguide is coupled to an adjacent waveguide with $\frac{\pi}{2}$ phase shift, and the remaining (determined by power transmission coefficient $t$) is transmitted through the input waveguide ($\kappa=t=$~0.5 in an ideal 50:50 DC). Throughout this process, the optical signal suffers from some optical loss based on the relationship $|\kappa|^2+|t|^2=\alpha_L$. Note that both $\kappa$ and $t$ are wavelength-dependent and they also depend on the waveguide width and thickness and the gap in DCs. The metal absorption loss ($\alpha_m$) is due to the absorption through metal planes of phase shifters in proximity to waveguides and it depends on the integration, material, and size of the metal planes \cite{ding2016broadband}. The waveguide propagation loss ($\alpha_p$) stems from the waveguide sidewall roughness and scattering loss \cite{Bahadori:16}. Considering optical losses $\alpha_L$, $\alpha_m$, and $\alpha_p$, a compact optical-loss-aware transfer-matrix model for the MZI in Fig.~\ref{MZI_schematic} can be defined as:
\begin{align*}
\begin{pmatrix}
O_{1} \\
O_{2}
\end{pmatrix}&= \begin{pmatrix}
T_{11} & T_{12} \\
T_{21} & T_{22}
\end{pmatrix}  \cdot 
\begin{pmatrix}
I_{1} \\
I_{2}
\end{pmatrix}
= T_{DC_{2}} \cdot T_{\theta} \cdot T_{DC_{1}} \cdot T_{\phi} \cdot \begin{pmatrix}
I_{1} \\
I_{2}
\end{pmatrix},\tag{1}\label{eq::MZI_trans_mat}\\
T_{DC_{2}}&=\begin{pmatrix}
\alpha_{L}\sqrt{1-\kappa_{2}} & \alpha_{L}j\sqrt{\kappa_{2}} \\
\alpha_{L}j\sqrt{\kappa_{2}} & \alpha_{L}\sqrt{1-\kappa_{2}}
\end{pmatrix},    
T_{\theta}=\begin{pmatrix}
\alpha_{p}l_{MZI}\alpha_{m}e^{j\theta} & 0 \\
0 & \alpha_{p}l_{MZI}
\end{pmatrix}\\
T_{DC_{1}}&=\begin{pmatrix}
\alpha_{L}\sqrt{1-\kappa_{1}} & \alpha_{L}j\sqrt{\kappa_{1}} \\
\alpha_{L}j\sqrt{\kappa_{1}} & \alpha_{L}\sqrt{1-\kappa_{1}}
\end{pmatrix},
T_{\phi}=\begin{pmatrix}
\alpha_{m}e^{j\phi} & 0 \\
0 & 1
\end{pmatrix}.    
\end{align*}
Here, $\kappa_{1}$ and $\kappa_{2}$ are the coupling coefficients in DC$_1$  and DC$_2$, respectively. Without loss of generality and in the absence of process variations, we assume $\kappa_1=\kappa_2$ ($\kappa_{1/2}=$~0.5 in 3-dB DCs). Moreover, $\alpha_{p}l_{MZI}$ is the MZI propagation loss where $\alpha_p$ is the waveguide propagation loss and $l_{MZI}$ is the MZI length (see Fig.~\ref{MZI_schematic} and Table~\ref{table_1}).

\begin{figure*}[ht]
  \centering
  \subfigure[MZI insertion loss]{
\includegraphics[width=.22\textwidth]{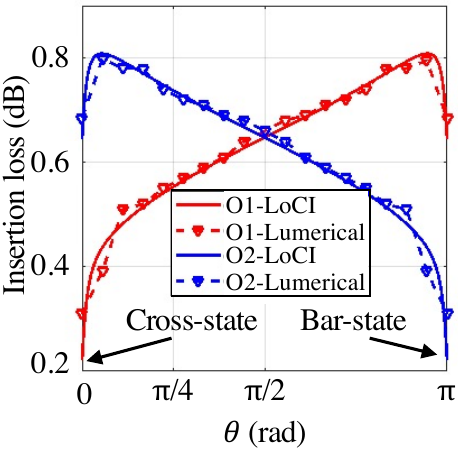}\label{MZI_loss_res}
}%
\subfigure[MZI crosstalk power]{
\includegraphics[width=.28\textwidth]{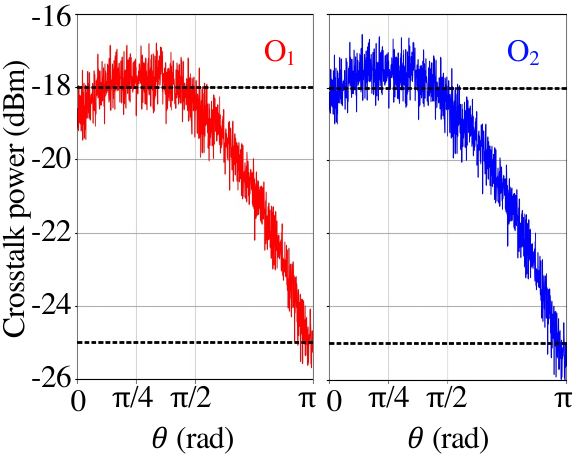}\label{MZI_xtalk_res}
}%
\subfigure[Single-layer OIU insertion loss]{
\includegraphics[width=.21\textwidth]{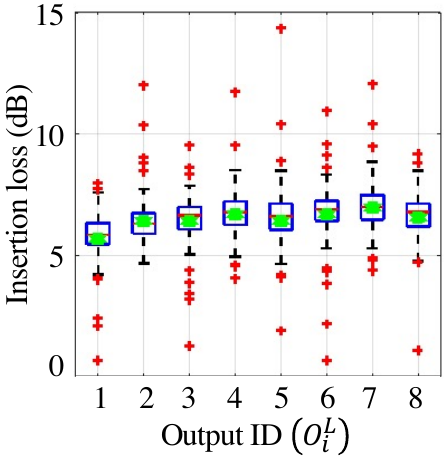}\label{layer_loss_res}
}%
\subfigure[Single-layer OIU crosstalk power]{
\includegraphics[width=.215\textwidth]{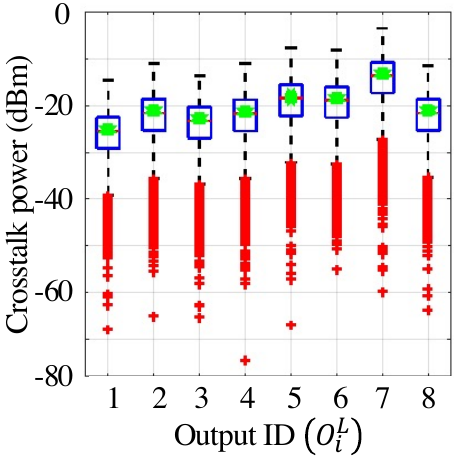}\label{layer_xtalk_res}
}%
\vspace{-0.15in}
  \caption{Insertion loss, (a), and crosstalk power, (b), at the output of the 2$\times$2 MZI in Fig.~\ref{MZI_schematic} simulated using the parameters listed in Table~\ref{table_1}. Boxplots for the insertion loss, (c), considering 100 random weight matrices and the coherent crosstalk noise power, (d), analyzed at each output of the OIU in a single layer ($M=$~1) with $N=$~8 (see Fig.~\ref{SP-NN_Arch}(a)). Green dots show the average results.}
 \vspace{-0.15in}
  \label{fig::device_results}
\end{figure*}

Optical crosstalk noise in an MZI can be analyzed by injecting an optical signal into a single input port at a time. That way, when $\theta=$~0 (Cross-state) or $\theta=\pi$ (Bar-state), the crosstalk coefficient can be captured on the opposite output port with destructive interference (see Fig. \ref{MZI_schematic}). For example, if $I_2$ is injecting, the Bar-state crosstalk coefficient ($X_B$) can be captured on $O_1$ when $\theta=\pi$ (hence $I_2\rightarrow O_2$), and the Cross-state crosstalk coefficient ($X_C$) can be captured on $O_2$ when $\theta=$~0 (hence $I_2\rightarrow O_1$, shown in Fig. \ref{MZI_schematic}). While this approach works for an MZI used as a switching cell---see Section \ref{sec::related_work}---it does not apply to the 2$\times$2 MZI multiplier in Fig.~\ref{MZI_schematic} whose $\theta$, which determines the MZI state, can be in the range 0$\leq\theta\leq\pi$ (note that $0\leq\phi\leq2\pi$ does not impact the MZI state). Consequently, there is no exact method to calculate the crosstalk coefficient on each output port because the MZI can be in an intermediate state (not only Bar- or Cross-state). To address this problem, we define a statistical model for the crosstalk coefficient ($X$) in the 2$\times$2 MZI multiplier in Fig.~\ref{MZI_schematic}. Considering the two known crosstalk coefficients $X_B$ and $X_C$, where typically $X_B\leq X_C$ \cite{Shoji:10Xtalk}, we analyze $X$ at an intermediate state determined by $\theta$ (and not by $\phi$) based on a Gaussian distribution with a $\theta$-dependent mean of $\mu(\theta)=\frac{X_B - X_C}{\pi}\theta+X_C$ and standard deviation of 0.05$\cdot\mu(\theta)$, considered here as an example. As a result and by employing \eqref{eq::MZI_trans_mat}, the coherent crosstalk noise on outputs $O_1$ and $O_2$ of the MZI in Fig.~\ref{MZI_schematic} can be modeled as (see also Fig. \ref{MZI_xtalk_res}):
\begin{equation*}\
\begin{pmatrix}
O_{1} \\
O_{2}
\end{pmatrix}=\begin{pmatrix}
(1-X)T_{11} & (1-X)T_{12} \\
(1-X)T_{21} & (1-X)T_{22}
\end{pmatrix}
\begin{pmatrix}
I_{1} \\
I_{2}
  \end{pmatrix}
  \\+\begin{pmatrix}
(X)T_{21} & (X)T_{22} \\
(X)T_{11} & (X)T_{12}
\end{pmatrix}
\begin{pmatrix}
I_{1} \\
I_{2}
\end{pmatrix}\tag{2}\label{eq::MZI_xtalk}. 
\end{equation*}
The proposed compact models in \eqref{eq::MZI_trans_mat} and \eqref{eq::MZI_xtalk} can be applied to any 2$\times$2 MZI structure in coherent SP-NNs.\vspace{-0.05in}  
\subsection{Layer-Level Compact Models}
As shown in Fig.~\ref{SP-NN_Arch}(a), we consider a generic coherent SP-NN model with $N_1$ inputs, $N_2$ outputs, and $M$ layers. Here, we assume $N=N_1=N_2$ for brevity. An optical signal in the input of a given layer goes through an array of cascaded MZIs in the OIU (see Fig.~\ref{SP-NN_Arch}(a)), where the number of MZIs depends on the OIU architecture \cite{Clements:16}. Note that $\theta$ and $\phi$ in each MZI, where $\theta$ determines the state and hence optical loss and crosstalk noise introduced in each MZI, depend on the weight parameters and can be determined using SP-NN training algorithms \cite{banerjee2021modeling}. The output of the OIU is connected to an optical-gain unit (OGU) that includes semiconductor optical amplifiers (SOAs) \cite{haq2020micro_SOA_Cband}. Last, the optical signal enters the nonlinear-activation unit (NAU), which can be implemented electronically \cite{SiPh_codesign}, optoelectronically \cite{PourFard:20}, or optically \cite{shen2017deep_nature}, each with different costs. Note that optical NAUs are still immature, and hence electronic and optoelectronic NAUs have been mostly employed in SP-NNs. Considering Fig.~\ref{SP-NN_Arch}(a), the insertion loss ($IL$) of layer $L_m$ in a coherent SP-NN can be systematically modeled as:
\begin{equation*}
IL_{m}= IL_{OIU} \cdot G \cdot IL_{NAU},\tag{3}\label{eq::layer_loss}  
\end{equation*}
where $IL_{OIU}$ is the insertion loss in the OIU that can be calculated based on \eqref{eq::MZI_trans_mat} for each MZI and it depends on the OIU architecture and $\theta$ phase settings in MZIs. Moreover, $G$ is the optical gain of the SOAs in the OGU and $IL_{NAU}$ is the insertion loss due to the NAU. In this paper, we consider the state-of-the-art SOA in \cite{haq2020micro_SOA_Cband} with $G=$17~dB, and we assume $IL_{NAU}=$~1~dB based on the optoelectronic NAU proposed in \cite{PourFard:20}, which realizes arbitrary activation functions. 

As optical signals traverse MZIs in the OIU in SP-NNs, some coherent crosstalk will be generated and propagated towards the output of each layer, and eventually the network. The coherent crosstalk power ($XP$) at the output of layer $L_m$ can be defined as:\vspace{-0.05in}
\begin{equation*}
XP_m = \sum_{j=1}^{N_{MZI}} \left(P \cdot X^{mj}_{MZI}(\rho) \cdot IL^{mj}_{OIU} \right) \cdot G \cdot IL_{NAU}.\tag{4}\label{eq::layer_xtalk}\vspace{-0.05in}    
\end{equation*}
In \eqref{eq::layer_xtalk}, $N_{MZI}$ is the total number of MZIs in the OIU in layer $L_m$ and $P$ is the input optical power. Moreover, $X^{mj}_{MZI}(\rho)$ can be calculated using \eqref{eq::MZI_xtalk} and is the coherent crosstalk on the output of layer $L_m$ originating in MZI $j$ in the OIU. Also, $\rho$ is the optical phase of the crosstalk signal. Similarly, $IL^{mj}_{OIU}$ is the insertion loss, which can be calculated using \eqref{eq::MZI_trans_mat}, experienced by $X^{mj}_{MZI}(\rho)$ as it traverses the OIU. Note that although SOAs can help improve the insertion loss in SP-NNs, the SOA optical gain will be also applied to the coherent crosstalk signals, thereby exacerbating coherent crosstalk noise in SP-NNs. By cascading the insertion loss and crosstalk models in \eqref{eq::layer_loss} and \eqref{eq::layer_xtalk} across multiple layers, we can analyze the network-level insertion loss and crosstalk power in coherent SP-NNs of any size.\vspace{-0.05in}

%% file: results.tex
\section{Simulation Results and Discussions}
We implemented the proposed analytical models in Section~\ref{sec::LoCI} along with a coherent SP-NN architecture model based on \cite{Clements:16} in MATLAB. For layer- and network-level analysis, we consider random weight matrices of different dimensions ($N=${8, 16, 32, and 64}), and use SVD to obtain $U$, $\Sigma$, and $V^{H}$ (see Fig.~\ref{SP-NN_Arch}(a)) for each layer with $M=${1,2, and 3}. We employ the algorithm proposed in \cite{Clements:16} to calculate the phase settings ($\theta$ and $\phi$) in the MZIs in the network (see our discussion in Section 2.2). Note that random weight matrices are only used in the layer- and network-level optical loss and crosstalk quantitative simulations, and the inferencing accuracy simulations are based on trained weight matrices (see Section \ref{sec::system-level-accuracy}). Table \ref{table_1} lists the device-level parameters used in the simulations.\vspace{-0.05in} 
\begin{figure*}[ht]
  \centering
  \subfigure[Average and worst-case insertion loss]{
\includegraphics[width=.32\textwidth]{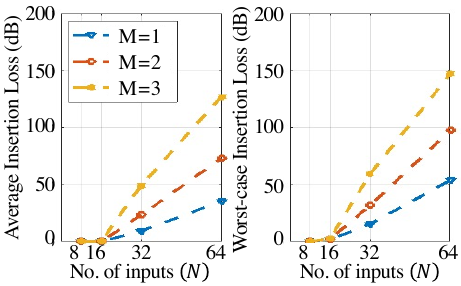}\label{fig::network_loss}
}%
\hspace{-0.3em}
\subfigure[Average and worst-case coherent crosstalk power]{
\includegraphics[width=.32\textwidth]{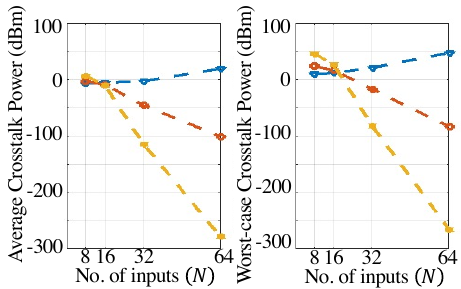}\label{fig::network_xtalk}
}%
\hspace{-0.3em}
\subfigure[Average and worst-case optical power penalty]{
\includegraphics[width=.32\textwidth]{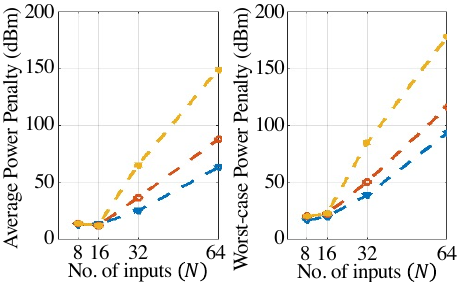}\label{fig::power_pen}
}
\vspace{-0.15in}
  \caption{The average and the worst-case insertion loss, coherent crosstalk power, and optical power penalty in coherent SP-NNs, based on the network in Fig.~\ref{SP-NN_Arch}(b) and with different numbers of inputs ($N$) and layers ($M$). The optical input power at layer one is 0~dBm in (a) and (b). Results are based on the parameters listed in Table~\ref{table_1}. Note that the average results are averaged among all the output ports in the network, and the worst-case results are based on the output port with the worst-case performance.}
 \vspace{-0.15in}
  \label{fig::network_results}
\end{figure*}

\subsection{Device-Level:
2$\times$2 MZI Multiplier}
Employing \eqref{eq::MZI_trans_mat} and \eqref{eq::MZI_xtalk} and the parameters listed in Table \ref{table_1}, Fig.~\ref{MZI_loss_res} and Fig.~\ref{MZI_xtalk_res} show the total insertion loss, which includes all the optical loss factors in \eqref{eq::MZI_trans_mat}, and crosstalk power at the output of the 2$\times$2 MZI in Fig. \ref{MZI_schematic}. The x-axis shows $0\leq\theta\leq\pi$,  which determines the MZI state ($\phi$ does not change the MZI state, but its loss is included). We used Lumerical \cite{Lumerical} to validate the results in Fig.~\ref{MZI_loss_res}. Note that Lumerical cannot analyze crosstalk in intermediate states, hence is not considered in Fig.~\ref{MZI_xtalk_res}. Observe that both the insertion loss and crosstalk noise power in the MZI change with the MZI state. The insertion loss on each output is $\approx$0.3--0.8~dB. Let us revisit Fig.~\ref{MZI_schematic}: compared to input $I_2$, the optical signal on $I_1$ experiences higher insertion loss because of $\alpha_{m}$ through $\phi$. Therefore, for example, the insertion loss is higher on $O_2$ ($O_1$) for the Cross-state (Bar-state). Note that the fluctuations in the crosstalk power in Fig. \ref{MZI_xtalk_res} are due to the Gaussian noise model defined for the MZI in Section~\ref{sec::device-compact-models}. The coherent crosstalk power in the MZI output changes between $\approx-$18~dBm and $\approx-$25~dBm, when the input power is 0~dBm.\vspace{-0.05in}
\begin{table}[t]
    \centering
    \caption{Device-level loss, crosstalk coefficient, power, and gain parameters considered in this paper (PhS: Phase shifter).}\vspace{-0.15in}
    \label{table_1}
\begin{tabular}{|c|c|c|c|} 
 \hline
 Par. & Definition & Value & Ref. \\ [0.5ex] 
 \hline\hline
 $X_B$ & Crosstalk in Bar-state &  -25 dB & \cite{Shoji:10Xtalk}\\ 
 \hline
$X_C$ & Crosstalk in Cross-state & -18 dB & \cite{Shoji:10Xtalk}\\
 \hline
 $l_{MZI}$ & MZI length  & 300 $\mu$m & \cite{Farhad_4by4}  \\
\hline
$\alpha_{m}$ & PhS (metal) absorption loss & 0.2 dB & \cite{ding2016broadband}\\
\hline
$\alpha_{p}$ & Propagation loss & 2 dB/cm & \cite{Bahadori:16}\\
 \hline
 $\alpha_{L}$ & Insertion loss of DC & 0.1 dB & \cite{Bahadori:16}\\
 \hline
 $L_{NAU}$ & NAU loss  & 1 dB & \cite{PourFard:20}\\
  \hline
 $G$ & SOA gain & ~17 dB (26.2 dBm) & \cite{haq2020micro_SOA_Cband}\\
 \hline
 $P$ & Input optical power & 0 dBm & -\\
 \hline
\end{tabular}
\vspace{-0.2in}
\end{table}

 \subsection{Layer-Level: Cascaded MZI Arrays (OIU)}
  We considered 100 random weight matrices with $N=$~8 (i.e., 64 MZIs in the OIU) and used \eqref{eq::layer_loss} to analyze the total insertion loss in one layer ($M=$~1). Results are shown in the boxplot (for 100 matrices) in Fig.~\ref{layer_loss_res}. Note that the insertion loss reported in Fig.~\ref{layer_loss_res} is analyzed at the output of the OIU and does not include the SOA gain ($G$) and NAU loss. Observe that the average and the worst-case insertion loss in the OIU of a fully connected layer with $N=$~8 are 6.5~dB and 14.4~dB, respectively. Similarly, using \eqref{eq::layer_xtalk} and a random weight matrix with $N=$~8, we analyze the coherent crosstalk power at the output of the OIU in a fully connected layer ($M=$~1). Considering \eqref{eq::layer_xtalk}, a coherent crosstalk signal arrives at an OIU output port with an optical phase $\rho$, where $0\leq\rho\leq2\pi$. Using a random uniform distribution between 0 and 2$\pi$, we assigned different optical phase angles, and repeated it 10000 times, to $\rho$ of the crosstalk signals at OIU outputs to statistically analyze the cumulative crosstalk signal interference at each output in the OIU. This approach is acceptable when optical signals traverse a large network of devices (e.g., in OIUs), and hence experience random phase shifts. Results are shown in the boxplot in Fig.~\ref{layer_xtalk_res}, where, similar to Fig.~\ref{layer_loss_res}, no SOA gain and NAU loss are considered. When $N=$~8 and $P=$~0~dBm, the average and the worst-case coherent crosstalk power at the output can be as high as $-$20~dBm and $-$3.8~dBm, respectively.\vspace{-0.05in}
  
\subsection{Network-Level: Coherent SP-NNs}
By extending the layer-level insertion loss and crosstalk models in \eqref{eq::layer_loss} and \eqref{eq::layer_xtalk} to full-network analysis, Fig.~\ref{fig::network_loss} and Fig.~\ref{fig::network_xtalk} show the average and the worst-case insertion loss and coherent crosstalk power, respectively, at the output of a coherent SP-NN as the number of inputs ($N$) and layers ($M$) are varied. In contrast to layer-level analysis studied in Section~4.2, the network-level results consider an SOA gain of 17~dB \cite{haq2020micro_SOA_Cband} and 1~dB loss per NAU \cite{PourFard:20} (see Table~\ref{table_1}) at the output of each layer. As shown in Fig.~\ref{fig::network_loss}, the insertion loss increases significantly as the number of inputs and layers increases. Even with a single layer ($M=$~1), the average (worst-case) insertion loss can be as high as 38.3~dB (54~dB) when $N=$~64. The drastically high insertion loss is due to the large number of cascaded MZIs in the OIUs (see Fig.~\ref{SP-NN_Arch}); this number is $MN(N-1)+MN$. 

Following the same coherent crosstalk noise analysis described in Section~4.2, Fig.~\ref{fig::network_xtalk} shows the average and the worst-case coherent crosstalk power in the SP-NN as the number of inputs and layers is increased. Note that the input optical power at the first layer is $P=$~0~dBm, and the crosstalk power results include the insertion loss---as well as the SOA gain---experienced by the crosstalk signals traversing the network. When $N$ and $M$ increase, the number of MZIs that generate coherent crosstalk towards the output ports increases as well, hence one would expect a higher crosstalk power at the output. However, crosstalk signals also experience a higher insertion loss as the network scales up (see Fig.~\ref{fig::network_loss}). Consequently, the coherent crosstalk power in the output can decrease when both $N$ and $M$ increase. As can be seen in Fig.~\ref{fig::network_xtalk}, when $M=$~1, the average (worst-case) coherent crosstalk power increases with $N$ and it can be as high as 19.6~dBm (48~dBm) when $N=$~64. Nevertheless, when both $N$ and $M$ increase, the severely higher resulting insertion loss diminishes the coherent crosstalk power in the output.\vspace{-0.05in}

\subsection{Optical Power Penalty in SP-NNs}
 \begin{figure*}[t]
  \centering
  \subfigure[Standalone impact of different sources of optical loss]{
\includegraphics[width=.3\textwidth]{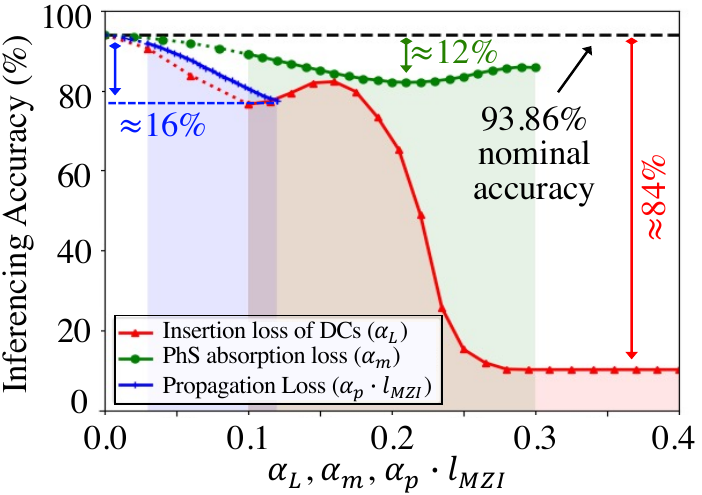}
}%
\hspace{-0.5em}
\subfigure[Simultaneous impact of different sources of optical loss]{
\includegraphics[width=.42\textwidth]{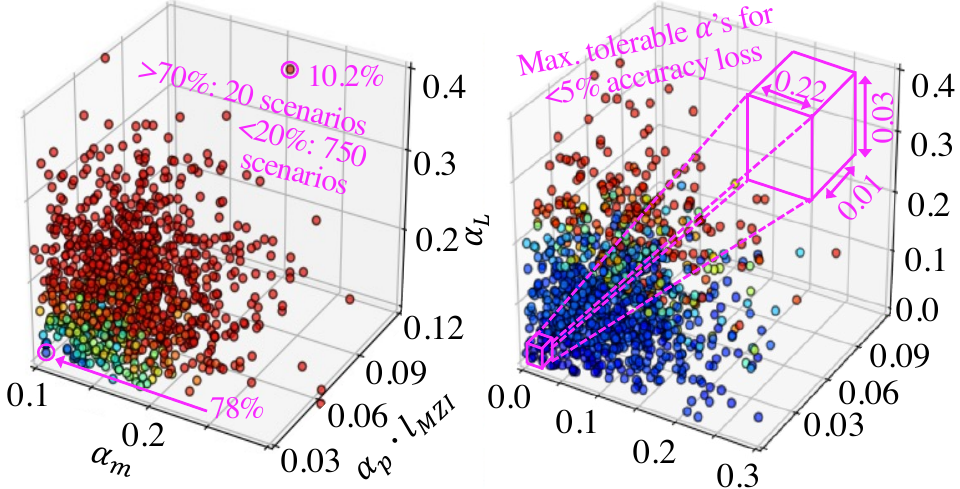}
}
\hspace{-1em}
\subfigure[Impact of loss and optical crosstalk]{
\includegraphics[width=.26\textwidth]{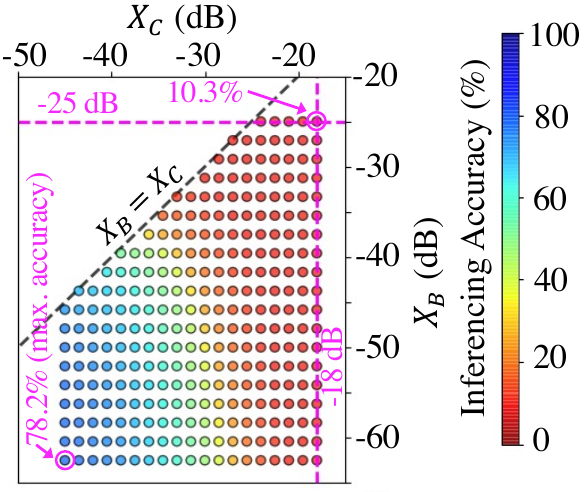}
}
\vspace{-0.15in}
  \caption{(a) SP-NN inferencing accuracy in the presence of DC insertion loss (red), absorption loss through phase shifter metal planes (green), and propagation loss in the MZI (blue)---see Table~\ref{table_1}. In each case, only one source of optical loss is considered at a time. The $\alpha$'s are expressed in dB and the shaded regions represent their respective expected range (see Section 2.3). The dotted section of each plot shows the accuracy loss for lower $\alpha$'s (outside their expected range). (b) Left: Inferencing accuracy when the loss parameters ($\alpha_{L}$, $\alpha_{m}$, and $\alpha_p\cdot l_{MZI}$) are simultaneously varied. Each of the 1000 points in the scatter plot represents an instance of the SP-NN where the $\alpha$'s are sampled from a half-normal distribution with mean, $\mu=$~their minimum expected value and standard deviation, $\sigma$, such that $3\sigma=$~their maximum expected value. Right: Inferencing accuracy when $\alpha$'s are sampled from a half normal distribution with mean, $\mu=$~0, $\sigma$, such that $3\sigma=$~their maximum expected value. (c) Inferencing accuracy in the presence of both optical loss and crosstalk noise for different values of $X_B$ and $X_C$ where $X_B\leq X\leq X_C$ (see Section~\ref{sec::device-compact-models}).}
 \vspace{-0.15in}
  \label{combo_system}
\end{figure*}
The optical loss and coherent crosstalk impose power penalty in SP-NNs. We study this issue by considering the input optical laser power ($P_{lsr}$) required at SP-NN input to compensate for the impact of loss and coherent crosstalk at the output (Fig.~\ref{SP-NN_Arch}(a)). For the network output $Y_y$ in a coherent SP-NN, the input optical laser power should satisfy the inequity $P_{lsr}\geq S_{PD} + IL^y + XP^y(\rho,P_{lsr})$. Here, $IL^y$ and $XP^y(\rho,P_{lsr})$ are the insertion loss (in dB) and coherent crosstalk power (in dBm), respectively, at the network output $Y_y$. They can be calculated using \eqref{eq::layer_loss} and \eqref{eq::layer_xtalk}, which include SOA gains in OGUs. Note that the total insertion loss for a victim signal at output $Y_y$ is determined by both $IL^y$ and the interference between the victim signal and the coherent crosstalk signal (determined by crosstalk signal phase $\rho$) at the same output, where the coherent crosstalk power also depends on $P_{lsr}$. Also, $S_{PD}$ is the sensitivity of the photodetector (in dBm) in electronic or optoelectronic NAUs \cite{PourFard:20}, taken to be $-$11.7~dBm in this paper \cite{9199100PD}. Considering the average and the worst-case insertion loss and crosstalk in Fig.~\ref{fig::network_loss} and Fig.~\ref{fig::network_xtalk}, Fig.~\ref{fig::power_pen} shows the average and the worst-case power penalty in coherent SP-NNs as $N$ and $M$ are increased, and without the OGU power penalty (26.2~dBm per SOA in Table~\ref{table_1}). Here, the interference between the victim signal and the coherent crosstalk signal at each output is explored statistically, and by considering both the average and the worst-case scenarios. On average, the optical power penalty to compensate for both insertion loss and coherent crosstalk is substantially high and easily exceeds 30~dBm when $N\geq$32, thereby considerably limiting SP-NN scalability. \vspace{-0.05in}

 \subsection{System-Level: Inferencing Accuracy}\label{sec::system-level-accuracy}

To analyze the system-level impact of optical loss and crosstalk, we consider a case study of an SP-NN with two hidden layers ($M=$~3) of 16 neurons each ($N=$~16), trained on the MNIST handwritten digit classification task. Each image in the MNIST dataset is converted to a complex feature vector of length 16 using the fast Fourier transform \cite{banerjee2021modeling}. The nominal test accuracy is 93.86\%. To analyze the effect of optical loss and crosstalk during inferencing, we integrated the MZI model in \eqref{eq::MZI_trans_mat} and \eqref{eq::MZI_xtalk} into our SP-NN model implementation.  




Based on our discussion in Section 2.3 and Table~\ref{table_1}, we consider $\alpha_{L}$, $\alpha_{m}$, and $\alpha_p$ within the range 0.1--0.4 dB \cite{Bahadori:16}, 0.1--0.3 dB \cite{ding2016broadband}, and 1--4 dB/cm \cite{Bahadori:16}, respectively. Considering an MZI of length $l_{MZI}=$~300~$\mu$m in~\cite{Farhad_4by4}, the propagation loss per MZI ($\alpha_p\cdot l_{MZI}$) is 0.03--0.12~dB. Fig. \ref{combo_system}(a) shows the inferencing accuracy of our example SP-NN when each of these $\alpha$'s are independently varied while the other $\alpha$'s are kept fixed at 0 dB and crosstalk is assumed to be absent. We observe that while the inferencing accuracy drops by up to 12\% and 16\% due to phase shifter metal absorption loss ($\alpha_m$) and the propagation loss ($\alpha_p\cdot l_{MZI}$), respectively, the impact of the DC insertion loss ($\alpha_L$) is significantly higher, and the accuracy can drop to $\approx$10\% for expected values of $\alpha_{L}$. Clearly, optical loss---and  DC insertion loss specifically---is catastrophic to network performance as also highlighted in Fig. \ref{combo_system}(b)-left, where we model an SP-NN under multiple simultaneous loss sources in the absence of crosstalk. Out of 1000 such random loss scenarios, we found that the SP-NN inferencing accuracy is less than 20\% in 750 scenarios and more than 70\% in only 20 scenarios. We found that, even when the $\alpha$'s are at their corresponding lowest expected values, the accuracy is only $\approx$~78\%. The maximum tolerable $\alpha$'s for which the accuracy loss is less than 5\% (in the absence of crosstalk) are shown in Fig. \ref{combo_system}(b)-right. To capture the impact of crosstalk on SP-NN inferencing accuracy, we determine crosstalk coefficient $X$ using a linear interpolation between the worst-case (Cross, $X_C=-$~18 dB) and the best-case (Bar, $X_B=-$~25 dB) crosstalk; see Section~\ref{sec::device-compact-models}. Fig. \ref{combo_system}(c) shows the inferencing accuracy in the presence of both optical loss and crosstalk, when $X_B\leq X\leq X_C$ and for different $X_B$ and $X_C$ and with $\alpha$'s set to their corresponding minimum expected values. When $X_C=-$18~dB and $X_B=-$25~dB, the accuracy drops to 10.3\%. We found that under optical crosstalk and average (or worst-case) loss, the accuracy remains at $\approx$~10\%. Even when $X_{B/C}$ decreases, the accuracy saturates at 78.2\% (lower left corner in Fig. \ref{combo_system}(c)). The results presented in this section motivate the need for SP-NN design exploration and optimization to mitigate optical loss and crosstalk. \vspace{-0.10in}

%% file: conclusion.tex
\section{Conclusion}
Optical loss and crosstalk noise in the underlying photonic devices in SP-NNs are critical roadblocks that limit SP-NN performance and scalability. In this paper, we have presented LoCI, the first modeling framework to characterize SP-NNs in the presence of optical loss and coherent crosstalk. We have analyzed the average and the worst-case insertion loss and coherent crosstalk noise, and their corresponding optical power penalty, in coherent SP-NNs while exploring inferencing accuracy drops in SP-NNs under such scenarios. Our results indicate the critical impact of optical loss and crosstalk noise in SP-NNs, resulting in significant power penalty and accuracy loss of 84\%. As SP-NNs are advanced to handle more complex problems, insights from this work can help photonic device engineers and SP-NN system architects to explore and optimize next-generation SP-NNs and evaluate SP-NN performance under critical optical loss and crosstalk noise.\vspace{-0.05in}
\section*{ACKNOWLEDGEMENTS}
This work was supported in part by the National Science
Foundation (NSF) under grant numbers CCF-1813370, CCF-
2006788, and CNS-2046226.